\newcommand\p{\pi}

\newcommand{\diracslash}[1]{#1\llap{/\kern2pt}}

\def\zbf#1{{\bf {#1}}}
\def\bearr{\begin{eqnarray}}
\def\eearr{\end{eqnarray}}
\def\zbf#1{{\bf {#1}}}

\newcommand{\be}{\begin{equation}}
\newcommand{\ee}{\end{equation}}
\newcommand{\bea}{\begin{eqnarray}}
\newcommand{\eea}{\end{eqnarray}}
\newcommand{\ba}[1]{\begin{array}{#1}}
\newcommand{\ea}{\end{array}}

\documentclass[preprint,secnumarabic,floats,amssymb,nobibnotes,nofootinbib,aps,prc,showpacs,tightenlines]{revtex4}
\usepackage{graphicx}
\usepackage{amsmath}
\usepackage{latexsym}
\usepackage{epsfig}
\usepackage{subfigure}
\begin{document}

\title{Warm natural inflation}
\author{
Hiranmaya Mishra$^a$, Subhendra Mohanty$^a$ and  Akhilesh Nautiyal$^b$ }
\affiliation{$^a$ Theory Division, Physical Research Laboratory,
Navrangpura, Ahmedabad 380 009, India\\
$^b$ Harish-Chandra Research Institute, Jhunsi, Allahabad, India }
\date{\today}

\begin{abstract}
In warm inflation models there is the requirement of generating
large dissipative couplings of the inflaton with radiation, while at the same
time, not de-stabilising the flatness of the inflaton potential due to radiative corrections.
One way to achieve this without fine tuning unrelated couplings is by supersymmetry.
In this paper we show that if the inflaton and other light fields are Pseudo-Nambu-Goldstone Bosons
then the radiative corrections to the potential are suppressed and the thermal corrections are small
as long as the temperature is below the symmetry breaking scale. In such models it is possible to fulfill
the contrary requirements of an inflaton potential
 which is stable under radiative corrections and the generation of
 a large dissipative coupling of the inflaton field with other light fields.  We construct a warm inflation model
which gives the observed CMB-anisotropy amplitude and spectral index where the symmetry breaking  is
at the GUT scale.
\end{abstract}


\maketitle

\section{introduction}
The cosmological theory of inflation \cite{Guth:1980zm} offers an  explanation of the
scale invariant horizon sized perturbations in the CMB observed by COBE \cite{Smoot:1992td} and WMAP \cite{Spergel:2003cb,Spergel:2006hy,Komatsu:2008hk,Komatsu:2010fb}.
A successful model of inflation requires the potential of the scalar field to be flat. A natural model for a flat potential is through Pseudo-Nambu-Goldstone bosons in a class of model called "natural inflation" \cite{Freese:1990rb,Adams:1992bn,Freese:2004un}. The flatness of the inflaton potential in Natural inflation models is ensured by the fact that the PNGB couplings with other fields are suppressed by the symmetry breaking scale $f$ and by making $f$ large enough one can control the quantum corrections to the inflaton potential. One major problem of the Natural inflation models is that the spectral index of the inflaton perturbations is related to the symmetry breaking scale as $n_s=1-M_P^2/(8 \pi f^2)$ and in order to be consistent with the WMAP \cite{Komatsu:2010fb}  measurement $n_s=0.963 \pm 0.014$, the symmetry breaking scale has to be close to the Planck scale \cite{Savage:2006tr}.
 In reference \cite{natural-gut}  it was pointed out that in warm inflation models \cite{Berera:1995ie, warm1, warm2} where there is a dissipative coupling between the inflaton and the radiation bath, one can have inflation with PNGB and the symmetry breaking scale could be lowered to the GUT scale $f\sim 10^{16}$GeV. In order to have a lower symmetry breaking scale in natural inflation models the dissipative coupling $\Gamma$ of inflaton with radiation must be much larger than $H$.

 Inflaton couplings with radiation fields through which one can
  generate large $\Gamma$ are also expected to spoil the flatness of the inflaton potential due to radiative corrections\cite{Yokoyama:1998ju,Berera:1998gx}. A solution for getting around the opposing constraints of obtaining
a large dissipative coefficient but with small corrections to the inflaton potential, without fine tuning unrelated couplings,
 was obtained by invoking supersymmetry
\cite{gamma-susy,Berera:2002sp}. Inflaton  was coupled to heavy catalyst
fields  with masses larger than the temperature of the Universe and these fields in turn were coupled to light fields.
The evolution of the inflaton would induce light particle production via the heavy catalyst fields. Since these heavy
catalyst fields are expected to populate mostly their ground state, the quantum corrections associated with them could be canceled in
super-symmetric models \cite{warm1} however the temperature dependent dissipative damping terms are not canceled.

In this paper we study warm inflation caused by
Pseudo-Nambu-Goldstone bosons as the inflaton. We
study the finte temperature corrections to the
potential. We show that the thermal corrections to the inflaton potential are small as long as the symmetry breaking
scale $f \gg T$.
We point out the possibility that if the light fields involved in the dissipative warm inflation
 are all PNGB's which arise in successive spontaneous symmetry breakings of the interactions of a heavy catalyst field then
 it is possible to generate a large dissipative couplings of the inflaton to radiation without destabilizing the inflaton
 potential. In the final section we show that taking the symmetry breaking near the GUT scale one can get warm inflation
 models with PNGB which satisfy all the observational constraints on the CMB-anisotropy spectrum from WMAP \cite{Komatsu:2010fb}.

\section{Warm inflation with PNGB}
Natural inflation models use the PNGB potential of the form which arise from an explicit symmetry breaking,
\be
V(\phi) = \Lambda(T)^4 \left(1- \cos\left(\frac{\phi}{f(T)}\right) \right).
\ee
Here we assume that the spontaneous symmetry breaking scale $f$  and the explicit symmetry breaking scale $\Lambda$ are
both dependent upon temperature through loop corrections. We will determine the finite temperature corrections to
$\Lambda$ and $f$ taking an example of PNGB in a $SU(N)$  technicolor theory where we assume fermions condensate are
formed giving rise to Goldstone bosons.
Consider a theory with N
flavors of fermions with an approximate chiral
$SU_L(N) \times SU_R(N)$
 symmetry. The axial part of the above
symmetry is broken spontaneously, when the fermion condensate
 $\langle\bar \psi \psi\rangle$
attains a nonzero vev. The low energy dynamics is then given in terms of the matrix field
$ U(x)=exp(2i\phi(x)/f)$
in terms of the $(N^2-1)$ Goldstone bosons $\phi(x)$. Here
$\phi(x)=\phi^a(x)T^a$, $a=1, N^2-1$ and $T^a$'s are the generators of
$SU(N)$.
We have shown in Appendix A that the temperature dependence of $\Lambda$ and $f$ are as follows ($\Lambda^4\equiv -\frac{2m_f}{N}\langle\bar\psi\psi\rangle$ of Eqn.~(\ref{pot})),
\be
\Lambda^4(T)
=\Lambda^4_0\,\left(1-\frac{2(N^2-1)}{N} \frac{T^2}{12f^2} \right),
\label{fft1}
\ee
and
\be
f(T) = f\left(1-N \frac{T^2}{24 f^2}\right)
\label{ft1}
\ee
to the leading  order in $T/f$.

The dynamics of inflaton field is governed by the equation
\be
\ddot{\phi}+(3H+\Gamma)\dot{\phi}+V^{\prime}(\phi,T)=0.
\label{inflaton}
\ee
Here over-dots represent derivative w.r.t $t$ and $\prime$ denotes differentiation with respect to $\phi$  and $\Gamma$ is the damping
term.
The total energy density and pressure of the system is given by
\bea
\rho&=&\frac{\dot{\phi}^2}{2}+V(\phi,T)+T s\, ,
\label{energy}\\
p&=&\frac{\dot{\phi}^2}{2}-V(\phi,T).
\label{presure}
\eea
Here $s$ is the entropy density of the system that is given by the derivative of the
potential with respect to temperature.
The Friedmann equation for expansion and the energy-momentum conservation equation are given by
\bea
H^2=\frac{8 \pi}{3 M_p^2}\rho\, ,
\label{friedmann}\\
\dot{\rho}+3 H(\rho+p)=0.
\label{energycon}
\eea
From above equations we get the energy conservation equation for the radiation as
\be
\dot{\rho}_r+4 H \rho_r=\Gamma \dot{\phi}^2.
\label{radiation}
\ee
During warm inflation $\rho_r$ remains nearly constant. Assuming a  slow roll of the inflaton
we neglect $\ddot{\phi}$ in the equation (\ref{inflaton}) and kinetic energy term in the
total energy density. During inflation the potential energy of the inflaton field is
dominant so we also can neglect the $T s$ term in the total energy density. So we get
\bea
\dot{\phi}&=&-\frac{V^{\prime}}{3H+\Gamma}\, ,\\
H^2&=&\frac{8 \pi}{3 M_p^2}V.
\label{friedmann1}
\eea
The slow role parameters are
defined as
\bea
\epsilon&=&\frac{M_p^2}{16 \pi}{\left(\frac{V^{\prime}}{V}\right)}^2\, ,\, \,  \eta=\frac{M_p^2}{8 \pi}\frac{V^{\prime\prime}}{V}\, ,\nonumber\\
\beta&=&\frac{M_p^2}{8 \pi}\frac{\Gamma^{\prime}V^{\prime}}{\Gamma V}\, ,\, \,
\delta=\frac{M_p^2}{8 \pi}\frac{T \frac{\partial V^{\prime}}{\partial T}}{V^{\prime}}\, ,\, \, \, c=\frac{T\frac{\partial \Gamma}{\partial T}}{\Gamma}
\eea
and the ratio between the thermal damping and the damping due to expansion is given by
\be
Q=\frac{\Gamma}{3 H}.
\ee
Here three extra slow roll parameters appear because of $\phi$ dependence of damping
term and temperature dependence of the potential and dissipation coefficient $\Gamma$. It was pointed out in \cite{Hall:2003zp} that these all parameters should be less than $1+Q$ to realize warm inflation. Latter, a detailed
stability analysis was made by Moss et al  \cite{Moss-Xiong} and they showed  that to have successful warm
inflation the parameter $\beta$ ($b$ in \cite{Moss-Xiong}) must be less than $\frac{Q}{1+Q}$ and $|c|<4$. This
implies that thermal corrections to the inflaton potential must be suppressed.
A detailed study of the perturbation spectra in warm inflation models has
been done by \cite{Hall:2003zp,Moss-Xiong,BasteroGil:2011xd} for constant as well as 
temperature dependent $\Gamma$.
However, we use the results of \cite{Hall:2003zp,Moss-Xiong} to calculate the
 power spectrum and spectral index for curvature perturbations, which are given as 
\bea
P_{\cal R}&=&\frac{\sqrt{\pi}}{2}\frac{H^3 T}{\dot\phi^2}\left(1+Q\right)^{\frac12},\label{power}\\
n_s&=&1+\frac{1}{\Delta}\left[-\frac{3\left(2 Q+2+5 Q c\right)\left(1+Q\right)}{Q}\delta-\frac{9 Q+17-5 c}{1+Q}\epsilon\right.
\nonumber\\
  &    &\left. - \frac{9 Q+1}{1+Q}\beta-\frac{3 Qc-6-6 Q+2c}{1+Q}\eta\right],
\label{ns}
\eea
 where $\Delta=4(1+Q)+(Q-1)c$.

 In the following sections we calculate the dissipation coefficient in a model where the inflaton and the light fields are PNGB fields which interact through a heavy catalyst field. We will compute the amplitude and spectral index of the curvature perturbation in  this model and compare with the WMAP observations to fix the scale of the symmetry breaking.

\section{Dissipation through Goldstone-bosons}
In our model, inflaton field $\phi$ transfers its energy to the field $\sigma$ via
intermediate fields $\chi$. We assume that both the inflaton and $\sigma$ fields are  PNGB with
spontaneous symmetry breaking scales $f$ and $f_\sigma$ respectively. We work in a high 
temperature regime where $f > T> m_\chi > m_\sigma, \,\, m_\phi$.

 The interaction of $\varphi$ with $\chi$ is described by the Lagrangian
\be
{\cal L}= \frac{1}{f^2}\left(\partial_\mu \varphi\right)\left(\partial^\mu \varphi\right)\chi^2
\label{lintchi}
\ee
and the interaction of $\chi$ with $\sigma$ fields is described by
\be
{\cal L}= \frac{1}{f_\sigma^2}\left(\partial_\mu \sigma\right)\left(\partial^\mu
 \sigma\right)\chi^2.
\label{lintsigma}
\ee
Such interactions are similar to Goldstone boson baryon interaction term in chiral 
perturbation theory \cite{Brown:1993yv}.
Integrating by parts, interaction~(\ref{lintchi}) can be re-expressed as
\be
{\cal L}=-\frac{1}{f^2}\left(\varphi\partial_\mu\partial^\mu \varphi\chi^2+
2\varphi\partial^\mu\varphi\chi\partial_\mu\chi\right)
\label{lint1}
\ee

Here $\varphi= \phi+\phi_1$, $\phi$ representing the zero mode
classical field. Hence the interaction term~(\ref{lint2}) with the classical background field 
can be expressed as
\be
{\cal L}= -\frac{1}{f^2}\left(\phi\ddot\phi \chi^2+2\phi\dot\phi\dot\chi\chi+
2\phi\partial^\mu\phi_1\chi\partial_\mu\chi\right)
\label{lint2}
\ee
In  slow-roll regime the $\ddot\phi$ term can be neglected. 
The dissipation coefficient $\Gamma$ can be derived from the effective equation of motion that
for the above interaction can be given as \cite{Berera:2002sp}
\be
\ddot\phi(t)+V^\prime(\phi)-\frac{2\dot\phi}{f^2}\langle\dot\chi\chi\rangle-
\frac{2}{f^2}\langle\partial^\mu\phi_1\chi\partial_\mu\chi\rangle=0, \label{effeom}
\ee
where $\langle...\rangle$ represents ensemble averages with respect to an equilibrium
(quantum or thermal) state. These field averages can be calculated using linear response theory
 and $\langle\dot\chi\chi\rangle$ can be written to the first order as
\be
\langle\dot\chi\chi\rangle=\langle\dot\chi\chi\rangle_0
-i\left(-\frac{2}{f^2}\right)\int d^4x^\prime\theta(t-t^\prime)
\left[\dot\phi(x^\prime)\phi(x^\prime)-
\dot\phi(x)\phi(x)\right]\langle\left[\dot\chi(x^\prime)\chi(x^\prime),\dot\chi(x)\chi(x)\right]\rangle.
\ee
Here $\langle\dot\chi\chi\rangle_0$ represents the correlation function evaluated  at initial 
time $t$. In slow-roll regime we can take $\dot\phi(t)$ as nearly constant so the effective equation
of motion (\ref{effeom}) becomes
\be
\ddot\phi(t)+V^\prime(\phi)-\frac{2\dot\phi}{f^2}\left[\langle\dot\chi\chi\rangle_0
-i\left(-\frac{2\dot\phi(t)}{f^2}\right)\int d^4x^\prime\theta(t-t^\prime)
\left(\phi(x^\prime)-\phi(x)\right)\langle\left[\dot\chi(x^\prime)\chi(x^\prime),\dot\chi(x)\chi(x)\right]\rangle\right]=0.
\ee

The non-local term $\left(\phi(x^\prime)-\phi(x)\right)$ in the integrand of the above equation 
can be simplified by Taylor expanding $\phi(x^\prime)$ and the effective equation of motion 
becomes
\be
\ddot\phi(t)+V^\prime(\phi)-\frac{2\dot\phi}{f^2}\left[\langle\dot\chi\chi\rangle_0
+i\left(\frac{2\dot\phi(t)^2}{f^2}\right)\int d^4x^\prime\theta(t-t^\prime)
\left(t^\prime-t\right)\langle\left[\dot\chi(x^\prime)\chi(x^\prime),\dot\chi(x)\chi(x)\right]\rangle\right]=0.\label{effeqm1}
\ee

\begin{figure}[ht]
\centering
\includegraphics[width=7.5cm,height=4cm]{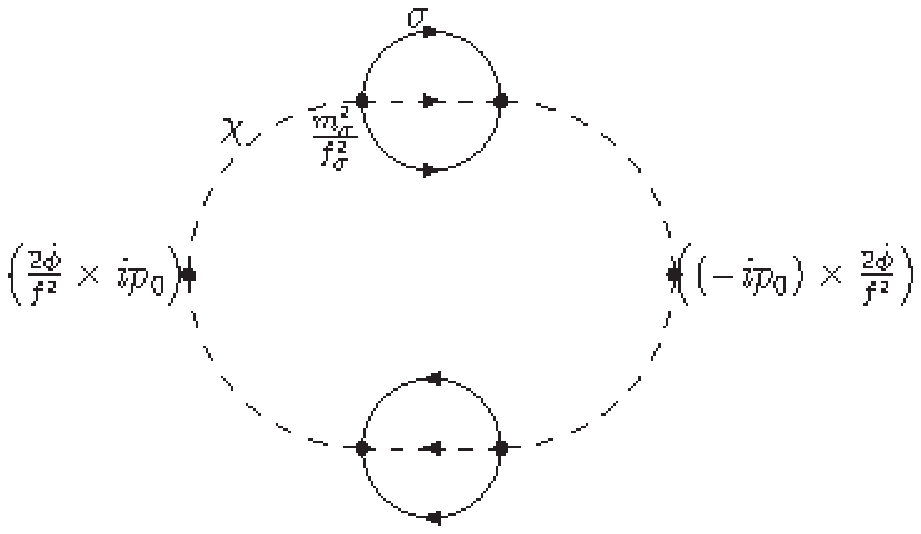}
\caption{ The self energy diagram of $\varphi$ which gives rise to the dissipation width $\Gamma$.}
\label{disspationdiag}
\end{figure}

We see that the contribution to the dissipation coefficient $\Gamma$ comes from the two coefficients of the 
$\dot\phi$ in Eq.~(\ref{effeqm1}). The first term is given by
\be
\Gamma_1=-\frac{2}{f^2}\langle\dot\chi\chi\rangle_0,
\ee 
while the second term will be
\be
 \Gamma_2=-i\left(\frac{4\dot\phi(t)^2}{f^4}\right)\int d^4x^\prime\theta(t-t^\prime)
\left(t^\prime-t\right)\langle\left[\dot\chi(x^\prime)\chi(x^\prime),\dot\chi(x)\chi(x)\right]\rangle\label{gammanew}
\ee
In terms of the spectral functions of the $\chi$ field, the first term becomes 
\cite{BasteroGil:2010pb}
\be
\Gamma_1=-\frac{2}{f^2}i\int \frac{d^4p}{(2\pi)^4} ip_0 \rho_\chi(p_0,\vec p)
n_\chi(p_0)\label{gammafirst}
\ee
where $n_\chi$ is the distribution function for the $\chi$ fields and $\rho_{\chi}$ is the spectral function 
 defined as
as
\be
\rho_{\chi}(p_0,p)=\frac{2Im \Sigma_{\chi}}{\left(p_0^2-\omega_p^2\right)^2+\left({Im\Sigma_{\chi}}\right)^2}
\label{spectral}
\ee
where $\omega_p=\sqrt{p^2+m_\chi^2}$. Using the relation $Im \Sigma_{\chi}=2 \omega_p\Gamma_{\chi}$ and substituting
(\ref{spectral}) in (\ref{gammafirst}) we get
\be
\Gamma_1=\frac{2}{f^2}\int \frac{d^3p}{(2\pi)^3}\frac{dp_0}{2\pi}p_0
\left[\frac{4\omega_p\Gamma_\chi}{\left(p_0^2-\omega_p^2\right)^2+(2\omega_p\gamma_\chi)^2}\right]n_\chi(p_0)
\ee
For a given temperature, the $p_0$ integral is dominated by the point $\omega_p$,
which lies close to the poles of the spectral function. So to  evaluate $p_0$ integral the
integrand can be expanded about $p_0=\omega_k$ and we obtain
\be
\Gamma_1=\frac{1}{f^2}\int \frac{d^3p}{(2\pi)^3}n_\chi(\omega_p)
\ee
that gives
\be
\Gamma_1\sim\frac{T^3}{f^2}.
\ee
This term is not sufficient for the strong dissipation as $\frac{T}{f}<<1$. To get large dissipation we will 
consider the second term (\ref{gammanew}) which can be represented 
diagrammatically as Fig.~\ref{disspationdiag}. As this is the main contribution to the dissipation coefficient
so we will leave the subscript $2$ in this term and denote it by $\Gamma$.

Again we can write $\Gamma_2$ in terms of spectral functions and we get \cite{BasteroGil:2010pb},
\be
\Gamma_2=\Gamma=\frac{2}{T}\frac{4}{f^4}\dot\phi^2\int \frac{d^4p}{(2\p)^4}p_0^2
\rho_\chi(p_0,\vec p)^2\times n_\chi\left(1+n_\chi\right)
\label{gamma}
\ee

Again we can follow the similar steps as earlier to calculate $\Gamma$. Here after evaluating the energy 
integral we get
\be
\Gamma=\frac{8}{ T} \frac{\dot\phi^2}{f^4}\int\frac{d^3p}{(2\pi)^3}\frac{1}{4\Gamma_{\chi}}
\times n_{\chi}(\omega_p)\left(1+n_{\chi}(\omega_p)\right).\label{gamma2}
\ee
The relaxation time $\tau_{\chi}$ for the $\chi$ can be calculated using
the interaction (\ref{lintsigma}). Since the $\sigma$ field is also a PNGB it has
 derivative coupling to the intermediate field $\chi$. We have used the method given in
 \cite{Hosoya:1983ke} to calculate the relaxation time. Here since the $\sigma$ field is in 
thermal bath, the contribution to the relaxation time of the $\chi$ field will come from the
thermal part of the free propagator of the internal lines. As it is evident from the finite 
temperature spectral function of the $\sigma$ field (see Eq.~(A10) of \cite{Hosoya:1983ke}), 
the derivative term in the interaction can be replaced by the mass of $\sigma$ field 
 using on-shell condition. Finally the relaxation time will be
\be
\tau_{\chi}^{-1}=\Gamma_{\chi}=\frac{3}{8\pi}\frac{m_\sigma^4}{f_\sigma^4}
\frac{T^2}{\omega_p}.
\ee
Putting this expression for $\Gamma_\chi$ in ~(\ref{gamma2}) and redefining the integration
variable $\beta\omega_p=x$  the dissipation coefficient will be
\be
\Gamma=\frac{8}{3\pi}\frac{\dot\phi^2}{f^4}\frac{f_\sigma^4}{m_\sigma^4}T
\int_{\beta m_\chi}^\infty x^2\sqrt{\left(x^2-(\beta m_\chi)^2\right)}
\frac{e^x}{\left(e^x-1\right)^2}dx.
\label{dissipation2}
\ee
Now in the regime $T>m_\chi$ the thermal integral can be evaluated numerically and it gives
a factor of $7$. Hence the dissipation coefficient will be
\be
\Gamma=\frac{56}{3\pi}\frac{\dot\phi^2}{f^4}\frac{f_\sigma^4}{m_\sigma^4}T
\label{dissipation3}
\ee 
We assume that $\phi$ and $\sigma$ are generated from successive symmetry breaking such that $f > f_\sigma$. The PNGB masses $m$ and $m_\sigma$ are related to the symmetry breaking scales $f$ and $f_\sigma$ respectively and some explicit symmetry breaking term in the Lagrangian like the fermions mass terms in a SU(N) chiral symmetry breaking model. We will assume for simplicity that the explicit symmetry breaking scale for both the PNGB are of the same magnitude $\Lambda$. Then from the Gell-Mann-Okubo-Rienner relation we have $m^2 f^2 = m_\sigma^2 f_\sigma^2= \Lambda^4$.
Using the GOR relation and denoting $c_\phi=(f_\sigma/f)$ the expression for $\Gamma$ in terms of $c_\phi$ can be written as
\be
\Gamma=\frac{56}{3\pi}\frac{\dot\phi^2c_\phi^8f^4}{\Lambda^8}T
\label{dissipation4}.
\ee

There will be thermal corrections to the effective potential of $\phi$ due to 
the $\chi$ loop that can be given as \cite{hep-ph/9211334} 
\be
V_{th}= \frac{m^2 \dot \phi^2}{8 \pi f^4} \left(\frac{T}{m_\chi}\right)\phi^2.
\ee
With the choice of the parameters required to satisfy CMB observations (see the next sections)
these corrections  will be of the order of
 $\frac{m^2 \dot \phi^2}{8\pi f^4}\left(\frac{T}{m_\chi}\right)\sim 10^{-8}\, GeV^2$ which are 
much smaller than the mass term of the inflaton i.e $\sim 10^{16}\, GeV^2$.
 
In the next section we will use this expression in the formula (\ref{power}) and (\ref{ns}) 
for the power spectrum amplitude and spectral index and fix the value of $f$ which gives the 
correct CMB spectrum.

\section{Observational constraints}
The PNGB potential parameters and warm inflation parameters can be constrained by using observational bounds on amplitude
and spectral index of primordial perturbations.

Using expression~(\ref{dissipation4}) for $\Gamma$, the inflaton equation of motion 
(\ref{inflaton}) can be rewritten as
\be
\ddot\phi+\left(3 H+\frac{56}{3\pi}\frac{\dot\phi^2c_\phi^8f^4}{\Lambda^8}T\right)\dot\phi+V^\prime(\phi)=0.
\label{eom2}
\ee
To determine $\dot\phi$ we assume  slow-roll approximation and we get
\be
\dot\phi=-\left(\frac{3\pi V^\prime\Lambda^8 }{56c_\phi^8f^4 T}\right)^{1/3}
\label{phidot}
\ee
We use this value of $\dot\phi$ in the rest of the numerical calculations and we get 
$\Gamma\sim 10^{12}$ GeV, which satisfies the conditions $\Gamma^2>>V''(\phi)\sim 10^{17}$GeV\textsuperscript{2} 
and $\Gamma>>H\sim 10^6$GeV. This justifies our assumptions used in Eq.~(\ref{phidot}).

Assuming the amount of radiation produced by dissipation is nearly equal to the radiation diluted due to expansion, the radiation density is given as
\be
4 H\rho_r=\Gamma\dot\phi^2.
\ee
Using $\rho_r=\frac{\pi^2}{30}g_\star T^4$ and (\ref{dissipation3}), (\ref{phidot}) the
temperature of the thermal bath is given as
\be
T=\left(\frac{15}{2\pi^2g_\star}\right)^{3/13}\left(\frac{3\pi}{56}\right)^{1/13}
\left(\frac{\Lambda^8\left(V'(\phi)\right)^4
}{H^3 c_\phi^8f^4}\right)^{1/13}.
\label{temp}
\ee

\begin{figure}
\begin{minipage}[t]{8cm}
\epsfig{file=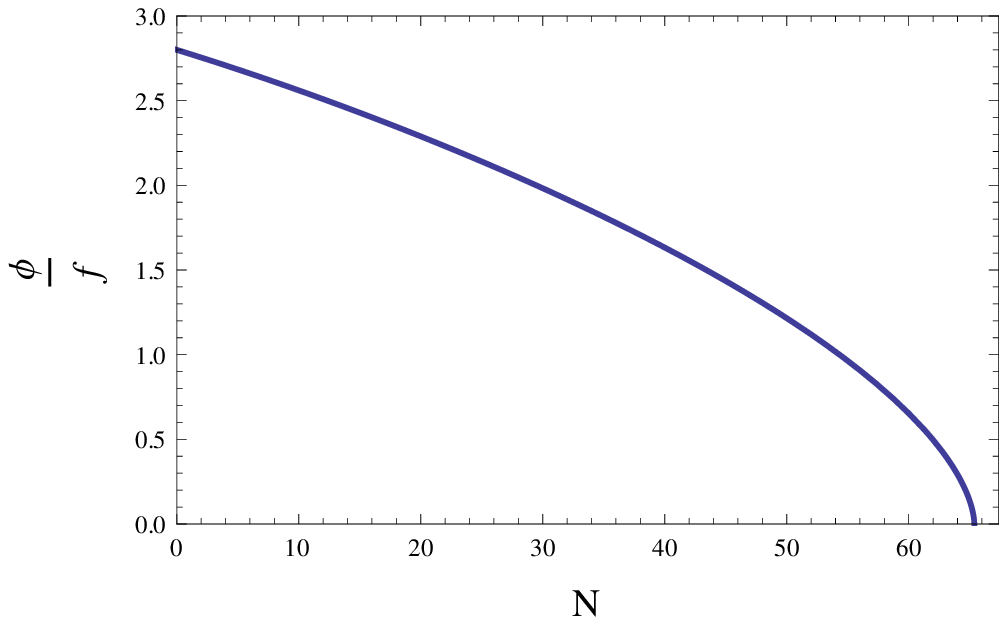,angle=0,width=7.5cm,height=6.6cm}
\caption{Evolution of inflation w.r.t efolds.}
\label{phin}
\end{minipage}
\hfill
\begin{minipage}[t]{7.5cm}
\epsfig{file=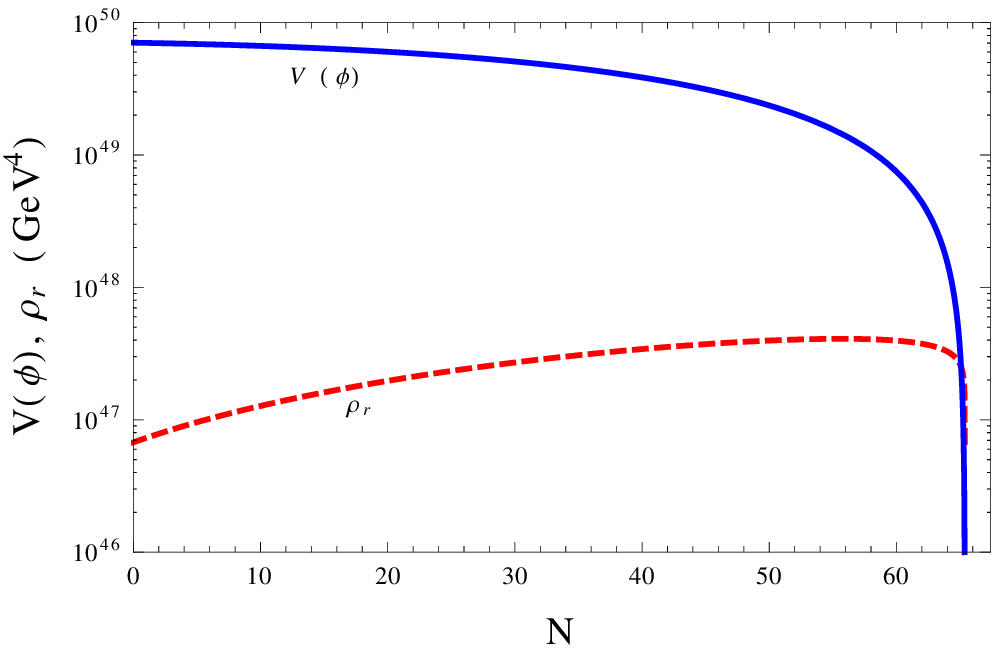,angle=0,width=7.5cm,height=6.6cm}
\caption{Radiation density and potential energy w.r.t efolds}
\label{vphirho}
\end{minipage}
\end{figure}

We have solved equation (\ref{eom2}) and 
$\dot H=-4\pi G \left(\dot\phi^2+\frac{4}{3}\rho_r\right)$
numerically using e-foldings $N=d\ln a$ as an independent variable. To solve these equation we take the values of the
 parameters that satisfy the observational constraints on $A_s$ and $n_s$. We take
 $c_\phi=0.145$,  $f= 1.29\times 10^{16}$GeV and
$\Lambda=2.45\times 10^{12}$GeV.
 The evolution of inflaton field  $\phi$, radiation density $\rho_r$ and potential 
$V(\phi)$ is shown in Fig.~\ref{phin} and \ref{vphirho}. As depicted in Fig.~\ref{vphirho}, 
inflation ends when radiation density becomes equal to the potential energy
of the inflaton field and we enter in a radiation dominated phase. With this choice of 
parameters, we can have sufficient number of e-foldings required to solve the horizon problem.

\begin{figure}[h]
\begin{center}
\epsfig{file=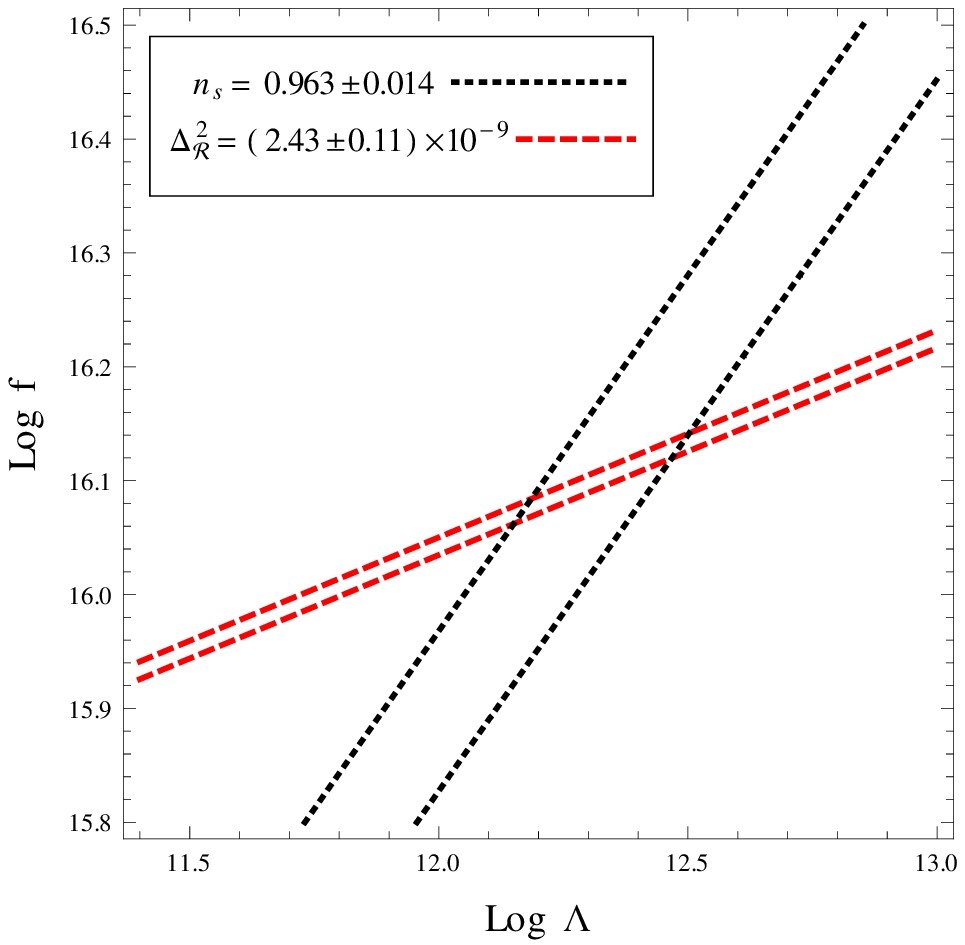,angle=0,width=10cm,height=8cm}
\end{center}
\caption{The allowed range of $f$(GeV) and $\Lambda$(GeV) using the WMAP 7 constraints on amplitude  ${\Delta_{\mathcal R}}^2$ and spectral index $n_s$ of curvature perturbations.}
\label{powerfig}
\end{figure}

We have three parameters $c_\phi$, $f$ and $\Lambda$ in this model that can be constrained
from observations. In Fig.\ref{powerfig}, we have plotted the power spectrum and spectral index by varying $f$ and $\Lambda$
 and using (\ref{dissipation4}) for $\Gamma$ and (\ref{temp}) for $T$. It is clear from the 
figure that the symmetry breaking scale is at the GUT scale to satisfy the observation constraints.
From this figure we can see that for $A_s=2.38\times 10^{-9}$ and $n_s=0.959$, which are 
within $1 \sigma$, $f=1.29\times 10^{16}$GeV and $\Lambda=2.45\times 10^{12}$GeV. 
So for $\phi=2.8\, f$,  we  get the values for other parameters using Eq~(\ref{dissipation4}), 
(\ref{temp}), (\ref{phidot}), (\ref{friedmann1}) and we find that $\Gamma=1.66\times 10^{12}$GeV, 
$T=2.09 \times 10^{11}$GeV and $H=1.99\times 10^6$GeV. The ratio of the dissipation coefficient 
and Hubble constant $Q\sim 10^5$ is quite large while the thermal corrections to the inflaton 
potential are suppressed as $T/f$.
\section{Conclusions}
We have shown that it is possible to achieve natural inflation at the GUT scale $f\sim 10^{16}$ GeV in warm inflation models if the
dissipative coupling of the inflatons is large $\Gamma \sim 10^{12}$GeV. We compute the thermal corrections to the
PNGB potential and show that the corrections are small as long as $T\ll f$. We also show that a large dissipative coupling can be achieved
if the both the inflaton and a light radiation field are PNGB's which arise in successive spontaneous symmetry breakings
and which couple to a heavy catalyst field. Thus PNGB model of
warm inflation is another class of models besides supersymmetry where one can naturally fulfill requirements of generating a
large dissipative coupling without destabilizing the inflaton potential by thermal and quantum corrections.

It has been shown recently \cite{Barnaby:2010vf,Barnaby:2011vw} that in a  model of inflation with PNGB
there is a non-gaussianity due to the coupling of the pseudo-scalar with gauge bosons.
In models such as ours where $f<<M_P$ the non-gaussianity expected to be large and
this could be observed in the forthcoming Planck \cite{:2006uk} observations.
\section{Acknowledgement}
 We thank the referee for useful suggestions.

\appendix*

\section{Thermal corrections to the PNGB potential}

Consider a theory with N
flavors of fermions with an approximate chiral
$SU_L(N) \times SU_R(N)$
 symmetry . The axial part of the above
symmetry is broken spontaneously, when the fermion condensate
 $\langle\bar \psi \psi\rangle$
attains a nonzero vev. The low energy dynamics is then given in terms of the matrix field
$ U(x)=exp(2i\phi(x)/f)$
in terms of the $(N^2-1)$ Goldstone bosons $\phi(x)$. Here
$\phi(x)=\phi^a(x)T^a$, $a=1, N^2-1$ and $T^a$'s are the generators of
$SU(N)$. Then the most general chirally invariant effective Lagrangian density
with minimal number of derivatives given as\cite{Scherer:2002tk}
\be
{\cal L}_{eff}=\frac{f^2}{4}Tr\left(\partial_\mu U\partial^\mu U^\dagger\right)
\label{lkin}
\ee
with $f$ as the corresponding 'pion' decay constant. An explicit symmetry
breaking gives rise to finite masses of the Goldstone 'pions' and is given by
\be
{\cal L}_{sb}=\frac{-\langle \bar \psi \psi \rangle }{2N}Tr\left(MU^\dagger+UM^\dagger \right)
\label{lsb}
\ee
where M is the fermion mass matrix which breaks the chiral symmetry explicitly.  In the following we shall also assume the
the  masses of the fermions to be the same $m_f$ for all the
flavors i.e $M=m_f I$ in which case the symmetry breaking term can be written explicitly in terms of the PNGB fields as
\be
{\cal L}_{sb}=\frac{- \langle \bar \psi \psi \rangle m_f}{N}
 Tr \cos\left(\frac{2 \phi^a T^a}{f}\right)
\label{pot}
\ee
where summation over all the flavors is understood for the condensate.

An explicit evaluation of the PNGB potential (\ref{pot}) up to  quadratic
 terms in the  fields  gives us the Gellmann-Oakes-Renner
(GOR) relation relating the PNGB mass $m$ with the explicit symmetry breaking scale $m_f$,
\be
m^2f^2=-\frac{2m_f}{N} \langle\bar \psi \psi \rangle .
\label{gor}
\ee

To calculate the the temperature dependence of the condensate, we use the
Feynman - Hellmann theorem, according to which the value of
the condensate $\langle\bar \psi \psi\rangle_T$ at finite temperature is related to the derivative
of the free energy density with respect to the symmetry breaking
parameter $m_f$ ,
 \be
 \langle\bar \psi \psi\rangle_T=
\langle\bar \psi \psi\rangle+ \frac{\partial}{\partial m_f}\tilde \Omega(T).
\label{psipsit}
\ee

 Assuming that the thermodynamic
potential is dominated
by the Goldstone modes,  the free energy difference
$\tilde \Omega(T)=\Omega(T)-\Omega(T=0)$ is given as
\be
\tilde\Omega(T)=\frac{(N^2-1)T}{(2\pi)^3}\int d\zbf k \ln(1-\exp^{-E/T})
\label{tomg}
\ee
where, $E=\sqrt{\zbf k^2+m^2}$ is the single PNGB energy. Next, we may
use the fact that $\frac{\partial m}{\partial m_f}=\frac{m}{2m_f}$ from
the  GOR relation Eq.(\ref{gor}) and, eliminate the condensate
$\langle\bar \psi\psi\rangle$ in favor of $m$ using the same equation to
obtain from Eq.(\ref{psipsit}) and Eq.(\ref{tomg}),
\be
\langle\bar \psi \psi\rangle_T
=\langle\bar \psi \psi\rangle_0\,\left(1-\frac{2(N^2-1)}{N}t_1\right),
\label{fft}                                                                     \ee
where,
\be
t_1=\frac{1}{f^2}\int \frac{d\zbf k}{(2\pi)^3}\frac{1}{2E}\frac{1}{\exp(-E/T)}
\simeq \frac{T^2}{12f^2}
\label{t1}
\ee
In the last step we have evaluated the integral in the
chiral limit $m << T$ \cite{Gasser:1986vb}.

Next, we proceed to calculate the leading temperature dependence of the masses
of the Pseudo Goldstone bosons. This can be calculated in the same line as
has been worked out in Ref.\cite{Chanfray:1996bd} for two flavor case using
Hartree approximation. The key quantity here is to calculate the pion self
energy in the medium. To lowest order, the interaction part is given by
collecting quartic order fields in the expansion of Eq.(\ref{lkin}) and
Eq.(\ref{lsb}) and takes the form
\bearr
{\cal L}_{int} &= &\frac{1}{f^2}\bigg[\frac{m^2}{3}Tr(\phi^4)+
4\bigg\{\frac{1}{4}Tr(\partial_\mu\phi^2)(\partial^\mu\phi^2)\nonumber\\
&-&\frac{1}{6}Tr(\partial_\mu\phi)(\partial^\mu\phi^3)
-\frac{1}{6}Tr(\partial_\mu\phi^3)(\partial^\mu\phi)\bigg\}\bigg].\\
\label{lint}
\eearr
\begin{figure}[ht]
\centering
\subfigure[]{
\includegraphics[width=7.5cm,height=4cm]{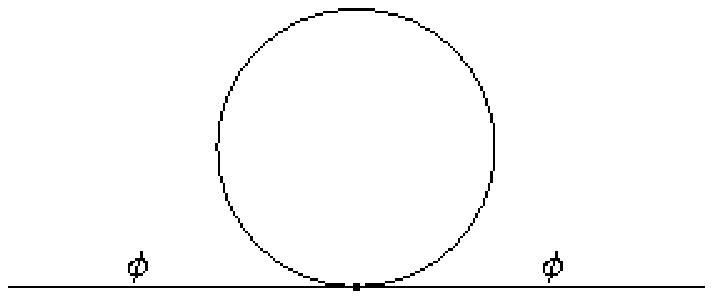}
\label{fig:tadpole1}
}
\subfigure[]{
\includegraphics[width=7.5cm,height=4cm]{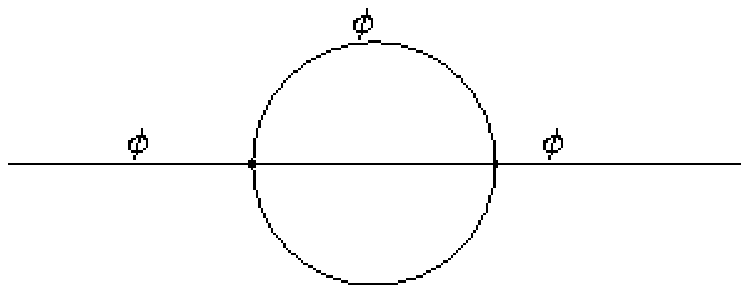}
\label{fig:tadpole2}
}
\caption{\em{ PNGB self energy}}
\label{fig:tadpole}
\end{figure}

With the above interaction term, the contribution to the self energy
is given by the tadpole diagram and the sun set diagram shown in
Fig.~\ref{fig:tadpole1} and \ref{fig:tadpole2}. The Hartree
approximation corresponds to evaluating these diagrams with the full
single particle Greens function, rather than the bare ones. Explicitly
within Hartree approximation, the PNGB self energy is then given
as \cite{Chanfray:1996bd}
\be
\Pi(\omega,\zbf k,T)=\left[-\frac{2 m^2}{3N}(2N^2-3)+(\omega^2-\zbf k^2+
\tilde m^2)\frac{2N}{3}\right] t
\label{piomgk}
\ee
In the above, the first term in RHS originates from the term
proportional to $m^2$ in the Lagrangian the interaction term in Eq.
(\ref{lint}). The other terms originate from the derivative couplings
in the Lagrangian Eq.(\ref{lint}) acting respectively on the external
PNGB propagator  ($\omega^2-\zbf k^2$) and on the PNGB loop ($\tilde m^2$).
 $\tilde m$ is the in medium PNGB mass.
Here we have used the trace relation for the $SU(N)$ generators
$Tr(T^aT^cT^cT^b)=(N^2-1)/(4N) \delta^{ab}$ and
$Tr(T^aT^cT^bT^c)=-1/(4N) \delta^{ab}$. further, we have here
included the  temperature dependent part of the propagator for the
PNGB field and have discarded the vacuum contribution.
The term $t$  is the same as in Eq.(\ref{t1})
except the change $E\rightarrow \tilde\omega=\sqrt{(\tilde m^2+\zbf k^2)}$,
denoting the in medium PNGB dispersion relation with the effective
mass $\tilde m$.
The effective mass
satisfies the equation
\be
\tilde m^2=m^2+\Pi(\tilde m,\zbf k=\zbf 0,T)
\ee
leading to
\be
\tilde m^2=
\frac{1-\frac{2}{3N}(2N^2-3) t_1}{1-\frac{4N}{3} t_1}m^2
\label{tmpi}
\ee
further noting the fact that the GOR relation is also valid at finite
 temperature\cite{Chanfray:1996bd}, we obtain the temperature dependence of the
 the decay constant, using Eq.(\ref{fft}) and Eq.(\ref{tmpi}),as
\be
f(T) = f(1-N \frac{T^2}{24 f^2})
\label{ft}
\ee
to the leading  order in $T/f$.

\end{document}